\documentstyle[12pt]{article}
\begin{document}
\baselineskip  0.7cm
\begin{center}
SUPERCONDUCTIVITY OF NEARLY 2-D SYSTEMS\\[0.3cm]
by\\
Nguyen Minh Khue \\
{\it Institute of Theoretical Physics\\
National Centre for Natural Science and Technology of Vietnam\\
P.O.Box 429, Hanoi 10000, Vietnam}\\

and \\
Le Ngoc Minh\\
{\it Hue University, Hue City, Vietnam}\\

and\\ Nguyen Tri Lan\\
{\it Institute of Theoretical Physics\\
National Centre for Natural Science and Technology of Vietnam\\
P.O.Box 429, Hanoi 10000, Vietnam}\\

and\\
Do Huu Nha\\
{\it National Pedagogic Institute of Vietnam\\
Dich vong, Tu liem, Hanoi, Vietnam}
\end{center}
\vspace{2cm}
\newpage
\begin{abstract}
Effect of geometry on the superconductivity is considered. It is shown that
the
for nearly two dimensional BCS systems the critical temperature is rapidly
increased with decreasing the thickness of the layer. The result is expected
to
be useful in the explanation of high-Tc conductivity of the ceramic compounds.
\end{abstract}

\newpage
\noindent
{\bf 1. Introduction}

In recent years finding mechanisms for high-Tc superconductivity is one of the
most interesting topic in Condensed matter Physics. Commonly it is believed
that the reasons leading to the high-Tc superconductivity of ceramic compounds
are the strong correlation between electrons or the two-dimensional (2-D)
character of the crystals or both of them.
In almost theories ideal 2-D systems have been considered [1].
In this work we consider
the superconductivity of nearly 2-D BCS systems and show that for such
systems the critical temperature $T_c$ is very
sensitive to the thickness $d$ of the layer, Relation between $T_c$ and $d$ is
established in section 3 basing on the model described in Section 2.\\[0.5cm]
{\bf 2. The Model}

Let us consider a nearly 2-D interacting  electron  system  with
thickness $d$. In
the coordinate system with z-axis perpendicular to the layer the system can be
described by the following Hamiltonian

\begin{equation}
H = \sum_i [-\frac{\hbar^2}{2m} \bigtriangledown_i^2 + v(\vec{r}_i)] +
\frac{1}{2} \sum_{i\ne j} V(\vec{r}_i-\vec{r}_j)
\end{equation}
where
\begin{equation}
v(\vec{r}) = \left\{ \begin{array}{ll}
o & \forall x,y \hspace{.5cm} {\rm and} \hspace{.5cm} -d/2 \le z \le d/2\\
\infty &\forall x,y \hspace{.5cm} {\rm and} \hspace{.5cm} z < -d/2
\hspace{.5cm} {\rm or} \hspace{.5cm} z>d/2
                   \end{array} \right.
\end{equation}

The first term in the r.h.s. of (1) describes non-interacting electrons, its
eigenfunctions and eigenvalues are
\begin{eqnarray}
\varphi_{n \vec{K}} &=& (dS)^{-1/2} sin(\frac{\pi n z}{d})
e^{i\vec{K}\vec{X}}\\
\in_{n \vec{K}} &=& \frac{(\pi \hbar n)^2}{2md^2} + \frac{\hbar^2\vec{K}^2}{2m}
,
\hspace{1cm} n = 1,2,3...
\end{eqnarray}
in which $\vec{K}$ and $\vec{X}$ are 2-D vectors : $ \vec{K} = (k_x,k_y),
\vec{X} = (x,y)$, and $S$ is the area of the system. In the following
the symbols $\in_n \equiv (\pi \hbar
n)^2/2md^2, \in_{\vec{K}} \equiv \hbar^2 \vec{K}^2/2m$ will be used
and $\in_{\vec K}$
will be measured from $\in_{\vec K_F}$ with $\vec K_F$ being the Fermi
momentum. The structure of the spectrum is
illustrated in Fig.1. The occupation of single states depends on the
thickness,
some possibilities are shown in Fig.2.

Using (3) as basic functions for the second quantization and remaining only
matrix elements describing the scattering between electrons of opposite
momentum and spin we have
\begin{equation}
H = \sum_{nK} \in_{nK}a_{nK}^+a_{nK} +
 \frac{1}{2} \sum_{nK,n'K'} V_{nK,n'K'}
a_{n,K}^+a_{n,-K}^+a_{n',-K'}a_{n',K'}
\end{equation}
in which $a_{n,K}^+$ and $a_{n,K}$ are the creation and annihilation operators
for electron on branch $n$ with momentum  $\vec{K}$ and spin up and
$a_{n,-K}^+$ and $a_{n,K}$ are that for electron on branch $n$ with momentum
$-\vec{K}$ and spin down.

As in BCS theory the sum in the second term in the r.h.s. of (5) will be
restricted by the condition $\vert \in_{\vec{K}} \vert \le \hbar \omega_D$
with $\omega_D $ being the Debye frequency.\\[0.5cm]
{\bf 3. Superconductivity}

Following BCS we consider the pairing between electrons with opposite
momentum and spin with energy $\vert \in_{\vec{K}}\vert \le \hbar
\omega_D$, i.e, we chose the trial function in the following form

\begin{equation}
\vert \psi> = \prod_{n,K}(u_{n,K} + v_{n,K} a_{n,K}^+ a_{n,-K}^+) \vert O>
\end{equation}
with the conditions
\begin{equation}
u_{n,K}^2 + v_{n,K}^2 = 1
\end{equation}
\begin{equation}
u_{n,-K} = u_{n,K} , \qquad v_{n,-K} = -v_{n,K}
\end{equation}

Note that in state $\vert \psi>$ there are only pairs of electrons on
the same branch of spectrum since the two conditions $\vert
\in_{\vec{K}}\vert \le \hbar \omega_D $ and $ \vert \in_{-\vec{K}}
\vert \le \hbar \omega_D$ can not be satisfied simultaneously for
different branches.

Carrying the same calculations as in the traditional BCS theory [2]
we obtain the following system of equations for the gaps
\begin{equation}
\Delta_{n,K} = -\frac{1}{2} \sum_{n',K'} V_{nK,n'K'}
\frac{\Delta_{n',K'}}{E_{n',K'}} th \frac{1}{2} \beta E_{n'K'}
\end{equation}
in which $E_{nK} = (\in_{nK}^2 + \Delta_{nK}^2)^{1/2}$, $\beta =
1/K_BT$, $K_B$ is the Boltzman constant and $T$ is the temperature.

After investigating Eq.(9)  we  note  that  for  small  $d$  the
interaction
$V(\vec r_i - \vec r_j)$ is weakly dependent  on  $z_i-z_j$
therefore  we may assume  that $V(\vec r_i - \vec r_j) = V(\vec X_i -
\vec X_j)$  and therefore $V_{n,K,  n'K'}  = \delta_{nn'}V_{K,K'}$.
We consider two cases.

(i) For the case shown in Fig.2a we may assume
\begin{equation}
V_{K,K'} = \left\{ \begin{array}{ll}
    & -(V/S) \ \ {\rm if} \ \ \vert \in_{K} \vert \ \ {\rm and} \ \ \vert
\in_{K'} \vert \le \hbar \omega_D \ \
 {\rm and \ \ spins \ \ in \ \ states} \  K \ {\rm and} \  K' \\
& {\rm are \ \ the \ \ same \ \ and \  \ } K {\rm \ and \ \ }K' {\rm \
belong \ \ to \ \ branch} \ n=1 \\[0.3cm]
    & 0 \hspace{1cm} {\rm otherwise}
    \end{array} \right.
\end{equation}

Substituting (10) into Eq.(9) we obtain the following equation for
the temperature $\beta_{1c}$ for which $\Delta_1(\beta_{1c})=0$.
\begin{equation}
\int_0^{\hbar \omega_D} \frac{d \xi}{\xi + \in_1} th \frac{1}{2}
\beta_{1c} (\xi + \in_1) = \frac{1}{N(0)V}
\end{equation}
in which $N(0)$ is the density of states corresponding to
$\in_{\vec{K}}$ at the Fermi level.

(ii) For the case shown in Fig.2b we may assume
\begin{equation}
V_{K,K'} = \left\{ \begin{array}{ll}
    & -(V/S) \ \ {\rm if} \ \ \vert \in_{K} \vert \ \ {\rm and} \ \ \vert
\in_{K'} \vert \le \hbar \omega_D \ \
 {\rm and \ \ spins \ \ in \ \ states} \ K \ {\rm and} \ \ K'\\
& {\rm are \  the \  same \  and \  both \ \ }K{\rm \ \ and \
}K'{\rm \ \ belong \ to \ branch \ }n=1{\rm  \ \ or \ \ }n= 2\\[0.3cm]
    & 0 \hspace{1cm} {\rm otherwise}
    \end{array} \right.
\end{equation}

Substituting (12) into (9) we see $\Delta_{nK}=0$ for $n \ne 1 , 2$
and $ \Delta_{nK}=0$ if $\vert
\in_{K} \vert > \hbar \omega_D$, \ \ $\Delta_{nK} = \Delta_n$ if $\vert
\in_{K} \vert \le \hbar \omega_D$ for $n=1,2$ and $\Delta_n$ is the solution of
the following equation

\begin{equation}
[1-\frac{V}{2S} \sum_K \frac{1}{E_{nK}}th \frac{1}{2}\beta E_{nK}]
\Delta_n = 0.
\end{equation}

The  nontrivial solution of Eqs.(13) can be found by solving the
following equation.
\begin{equation}
\frac{V}{2S} \sum_K \frac{1}{E_{nK}}th \frac{1}{2} \beta
E_{nK} = 1.
\end{equation}

Setting $\Delta_n(\beta_{nc}) =  0  (n=1,2) $ in (14)
we obtain the following equation
\begin{equation}
\int_0^{\hbar \omega_D} \frac{d \xi}{\xi+\in_n} th
\frac{1}{2} \beta_{nc} (\xi + \in_n) = \frac{1}{N(0)V}, \qquad (n = 1,2)
\end{equation}

Eqs.(11) and (15) give us
\begin{equation}
K_BT_{nc} = 1,14 (\hbar \omega_D + (n \pi
\hbar)^2/2md^2)e^{-1/N(0)V}, \qquad (n = 1,2)
\end{equation}

 By the definition the critical temperature
$T_c=T_{1c}$ since $T_{1c}<T_{2c}$

Formula (16) shows that the critical temperature is increased with
decreasing the thickness and we hope that it should be useful in the
explanation of the high-Tc superconductivity of the ceramic
compounds.\\[0.5cm]
{\bf ACKNOWLEDGEMENTS}

One of the  authors  (NMK)  should  like  to  express  his  deep
gratitude to Prof. A.Zawadowski, Prof. J.S\'olyom, Dr. P.Fazekas and Dr.
F.Woynarovich for permanent paying attention to him.

The work has been completed with the financial support from Vietnam
National Research Program in Natural Sciences.\\[0.5cm]
{\bf REFERENCES}
\begin{enumerate}
\item  A.P. Balachandran, E.Ercolessi, G.Morandi and Ajit Mohan
Srivastava., {\it The Hubbard Model and Anyon Superconductivity},
SU-4228-431, TPI-MINN-90/26-T May 1990.
\item  J.R.Schrieffer, {\it Theory of Superconductivity}, Benjamin, 1964.
\end{enumerate}
\newpage
\noindent
{\bf FIGURE  CAPTIONS}

Fig.1. The structure of the spectrum.

Fig.2. Some possibilities of the occupation of single particle states.
\end{document}